\documentclass[preprint,aps]{revtex4}
\begin{document}
\title{Phase transitions in Lu$_2$Ir$_3$Si$_5$}
\author{Yogesh Singh$^1$,  Dilip Pal$^1$, S. Ramakrishnan$^1$, A. M. Awasthi$^2$ and S.K. Malik$^1$}
\address{$^1$Tata Institute Of Fundamental Research,Bombay-400005, India\\
$^2$IUC-DAEF, Indore-452017, India.}
\begin{abstract}
\noindent
We report the results of our investigations on a polycrystalline sample of Lu$_2$Ir$_3$Si$_5$ 
which crystallizes in the U$_2$Co$_3$Si$_5$ type structure (Ibam). These investigations comprise powder X-ray diffraction, magnetic susceptibility, electrical resistivity and high temperature (120-300~K) heat capacity studies. 
Our results reveal that the sample undergoes a superconducting transition below 3.5~K. It also undergoes a first order phase transition between 150-250~K as revealed by an upturn in the resistivity, a diasmagnetic drop in the magnetic susceptibility and a large anomaly (20-30~J/mol~K) in the specific heat data. We observe a huge thermal hysteresis of almost 45~K between the cooling and warming data across this high temperature transition in all our measurements. Low temperature X-ray diffraction measurements at 87~K reveals that the compound undergoes a structural change at the high temperature transition. Resistivity data taken in repeated cooling and warming cycles indicate that at the high temperature transition, the system goes into a highly metastable state and successive heating/cooling curves are found to lie above the previous one and the resistance keeps increasing with every thermal cycle. 
The room temperature resistance of a thermaly cycled piece of the sample decays exponentialy  with time with a decay time constant estimated to be about 10$^4$~secs. 
The anomaly (upturn) in the resistivity and the large drop (almost 45\%) in the susceptibility across the high temperature transition  suggest that the observed structural change is accompanied or induced by an electronic transition.\\ 
\vskip 1truecm 
\noindent 
Ms number ~~~~~~~~~~~~PACS number:~71.30.+h, 61.50.Ks, 61.10.Nz, 65.40.Ba\\
\end{abstract}
\maketitle
\newpage
\section{Introduction}
\label{sec:INTRO}
\noindent
Rare earth ternary silicides, which form in a variety of crystal structures, have led
to a large number of studies due to their unusual ground states \cite{r1,r2}. Depending on the compound,
one has observed superconductivity \cite{r3,r4}, coexistence of magnetism and superconductivity \cite{r5}, 
reentrant superconductivity \cite{r6,r7} and magnetic ordering in the heavy electron state \cite{r8,r9}. 
As a part of our continuing studies of the magnetic, electronic and transport properties of ternary rare-earth (R) intermetallic compounds of the type R$_2$T$_3$X$_5$, where T is a transition metal and X is  an s-p element, we have recently become interested  in the compounds of the series R$_2$Ir$_3$Si$_5$ (R=La-Lu) since the isostructural compounds belonging to R$_2$Rh$_3$Si$_5$ \cite{r10} series exhibit unusual superconducting and exotic magnetic
properties at low temperatures. Earlier studies \cite{r11,r12} established that Ce$_2$Ir$_3$Si$_5$ is non-magnetic
presumably due to the large Kondo temperature which effectively screens out the 4f moment of Ce. Recently we have reported on the low temperature properties of the compounds of the series R$_2$Ir$_3$Si$_5$ (R=La-Tm) \cite{r13}. To the best of our knowledge, investigations on Lu$_2$Ir$_3$Si$_5$ have not been made prior to this study.
In this paper we report a comprehensive study of the structure, electrical resistivity, magnetic susceptibility and heat capacity of the non-magnetic compound Lu$_2$Ir$_3$Si$_5$. The susceptibility and resistivity measurements indicate  structural and  CDW like transitions at high temperature followed by a superconducting transition at low temperature. Such unusual properties have been reported earlier for tetragonal (P4/mbm) Lu$_5$Ir$_4$Si$_{10}$ \cite{r14,r15}, where
one has observed coexistence of novel charge density wave with superconductivity below 3.9 K. However, unlike Lu$_5$Ir$_4$Si$_{10}$, Lu$_2$Ir$_3$Si$_5$ undergoes a major structural transition to another orthorhombic structure with
doubling of the unit cell.  
\noindent
\section{EXPERIMENTAL DETAILS}
\label{sec:EXPT}
\noindent
A polycrystalline sample of Lu$_2$Ir$_3$Si$_5$ was prepared by the usual arc melting method. The constituent elements (Lu~-99.9\%, Ir~-~99.9\% ; Si -~99.999\%) were taken in stoichiometric proportion and arc-melted on a water-cooled copper hearth under Ti gettered high purity argon atmosphere. The resulting ingot was flipped over and remelted 6 times to promote homogenous mixing. The sample was wrapped in a zirconium foil, sealed in an evacuated quartz tube and annealed at 950~$^o$C for eight days. A piece of the sample was crushed into a fine powder for X-ray diffraction measurement using Cu K$\alpha$ radiation in a commercial diffractometer. The room temperature powder X-ray diffraction pattern of the sample could be indexed to the orthorhombic structure (U$_2$Co$_3$Si$_5$, space group Ibam) with no impurity lines. The structure of the unit cell of Lu$_2$Ir$_3$Si$_5$
is shown in  Fig. 1. At this point it is instructive to compare the structure of R$_2$Ir$_3$Si$_5$ with that of R$_2$Fe$_3$Si$_5$, the latter of which is known to display unusual superconducting and magnetic properties. Both series are derived from BaAl$_4$-type structure. R$_2$Fe$_3$Si$_5$ forms in the tetragonal structure in which two different sets of Fe sites form chains along [001] direction (Fe(2)) and isolated squares parallel to the basal plane (Fe(1)). The R$_2$Ir$_3$Si$_5$  forms in the orthorhombic structure where the arrangements of the [001] columns lead to a different coordination of the transition metal and of silicon. Here, a deformed square pyramid of silicon atoms surrounds two-thirds of the transition metal atoms and each of the remaining transition metal atoms is in the center of a silicon tetrahedron. The latter transition metal atom form chains along [001] direction. The rare-earth atoms in  R$_2$Ir$_3$Si$_5$ structure form a distorted square net with distances 3.9 to 4.2 $\AA$ within the layers and interlayer distances of 5.4-6.2~\AA. The nearest rare-earth distances in R$_2$Fe$_3$Si$_5$ is 3.7 \AA. The Rietveld fit \cite{r16} to the powder X-ray data of Lu$_2$Ir$_3$Si$_5$ was done and the parameters obtained from this fit are given in Table~I. The values for the lattice constants estimated from the fit are a= 9.91457(5) \AA, b= 11.28665(5)\AA and c= 5.72191(5)\AA. An earlier report \cite{r12}  has established that the  compound Ce$_2$Ir$_3$Si$_5$ crystallizes in the same structure.\\
A commercial Superconducting Quantum Interference Device (SQUID) magnetometer (MPMS5, Quantum Design, USA) was used to measure the temperature dependence of
the magnetic susceptibility $\chi$ in a field of 10~Oe for temperatures between 1.8 to 10~K to detect the superconducting transition and in a field of 0.1~T in the temperature range from 10 to 300~K. The resistivity was measured using a 
four-probe dc technique on a home built setup with contacts made using silver paint on a bar shaped sample 
1~mm thick, 10~mm long and 2 mm wide. The temperature was measured using a calibrated Si
diode (Lake~Shore~Inc., USA) sensor. The sample voltage was measured with a
nanovoltmeter (model 182, Keithley, USA)  with  a current of 5~mA using a
20~ppm stable (Hewlett Packard, USA) current  source. All the data were
collected using an IBM compatible PC/AT via IEEE-488 interface. For measuring resistance vs temperature for repeated cooling and warming cycles in the temperature range 1.8-300~K we used a commercial system (PPMS. Quantum Design) .
The heat capacity in zero field between 120 to 300~K was measured using a commercial DSC system. 
\section{RESULTS AND DISCUSSION}
\noindent
Panel (a) in Fig.~2 shows the temperature dependence of the electrical resistivity $\rho$ vs temperature from 1.8 to 300~K. 
The data were recorded while warming the sample from 1.8~K to 300~K. The inset in the same panel shows the low temperature behavior of $\rho$ from 1.8 to 10~K. From the inset one can clearly see that the resistivity sharply drops below 3.3~K. The resistive drop however, is not complete down to 1.8~K. Panel (b) of the same figure shows the temperature dependence of the low temperature FC (data was recorded when sample was cooled with field on) and ZFC (data were recorded while warming up in a field after the sample was cooled in zero field) susceptibility to highlight the existence of superconductivity in the sample. We can clearly see the abrupt diamagnetic signal below 3.3~K. However, the transition is again not complete. Diamagnetism in the FC curve shows the bulk nature of the superconductivity in this compound. The diamagnetic signal in the $\chi$ measurement together with the abrupt drop in the resistivity at around the same temperature, suggest the presence of  superconductivity in this sample although it seems to be dependent on the exact composition of the compound. Another sample from a different batch shows a drop to zero resistivity at 2.8 K but the high temperature phase transition was broader as compared to the previous sample. From panel (a) in Fig.~2, we can see that between 150 to 225~K the resistivity of the compound shows an upturn similar to the one usually observed in charge density wave or spin density wave (CDW/SDW) transition due to the opening up of a gap in the electronic density of states associated with these transitions. After reaching a maximum at about 154~K, the resistivity continues to show a metallic behavior down to the lowest temperatures before undergoing the superconducting transition. It is interesting to recall that we had recently observed a similar but much weaker anomaly in the resistivity of the compound Er$_2$Ir$_3$Si$_5$ below 150~K although there was no signature of the transition in the magnetic susceptibility for that sample. None of the other members of the series showed this anomaly \cite{r13}.\\ 
Panel (a) in Fig.~3 shows the resistivity for temperature scans while both cooling and warming the sample between 100-300~K at a rate of 2~K/minute. On cooling down from 300~K (the lower curve) we encounter the onset (upturn in resistance) of the transition at about 170~K. The resistance continues to rise until it reaches a maximum at almost 154~K after which it starts decreasing with decreasing temperature. We see that the transition is a rather broad one almost 20~K wide. While warming the sample (upper curve) we find that the onset/upturn occurs at almost 215~K. We have plotted d$\rho$/dT vs temperature in the inset of this figure. The sharp peaks in the d$\rho$/dT plot give us a better estimate of the transition temperatures which are 164~K while cooling from 300~K and 208~K while warming to 300~K. Thus, we can clearly see that there is a huge thermal hysteresis of almost 40$-$45~K between the up and down scans. This strongly suggests a first order transition for the system. Another feature of interest in the resistivity plot shown in the panel (a) of Fig.~3 is that after the transition, the warming curve lies above the cooling curve and does not come down and meet the cooling curve for temperatures beyond the transition. We have taken repeated cooling and warming measurements continuously for many cycles and find that the resistance always keeps increasing with each thermal cycle. We have shown this in panel (b) of Fig.~3 for two down and two up scans taken one after the other in the sequence down-up-down-up. It is clearly seen that between cooling down from and warming up to 300~K, the resistance value has increased in both cycles.\\
We have done measurements (not shown here) for 7 cycles {\sl i.e.} we start from 300~K and measure down to 5~K and then we measure while warming up to 300~K again (we call this sequence one cycle) and repeat this for upto 6 cycles. For each cycle, we see that the data point at 300~K forms a sort of ladder, which keeps climbing up with each thermal cycle. 
\par
In order to understand the nature of this phase transition, we have carried out low temperature 
powder X-ray diffraction at 87~K, which along with the data at 300~K is shown Fig.~4. It is evident from this figure that the sample undergoes a structural transition below the high temperature phase transition. Preliminary analysis suggests that a doubling of the
unit cell of Lu$_2$Ir$_3$Si$_5$ could account for this change. However, single crystal study is required to establish this conjecture.
We have carried out X-ray measurements on a thermally cycled piece of the sample to see whether some part of the low temperature (high resistance) phase remains when we return from low temperatures to 300 K thus causing the resistance to go up with every thermal cycle. However, that is not the case and the X-ray matches the room temperature X-ray for the virgin (not subjected to any thermal cycling) sample. It is possible that there is a large volume change across the high temperature transition and this causes micro-cracks to appear inside the sample and this in turn could cause the resistance to go up with every thermal cycle as we would be increasing the number of such cracks with each cycle. However, a 4-probe measurement of the room temperature resistance as a function of the distance between the voltage leads shows a linear behavior suggesting that no large cracks are appearing on thermal cycling. 
Interestingly we find that the resistance decays with time if left at 300~K after it has been cycled several times between 
5~K and 300~K. In Fig.~5 we show the resistance as a function of time measured for 24 hours when the sample is left at 300~K after being subjected to 7 thermal cycles between 5 and 300~K. Also plotted in the same figure is a fit to an exponentially decaying function of the form $\rho_0~e^{-t/\tau}$. The estimated time constant $\tau$ comes out to be 3.4$\times~10^4$~secs. The decay of the resistance after being subjected to several thermal cycles also contradicts the notion of the resistance increasing due to the cracks developing in the samples as the cracks would not anneal with time causing the resistance to decay.\\
The huge thermal hysteresis in the cooling and warming scans is also seen in the susceptibility $\chi$ vs. temperature plot shown in the panel (a) of the Fig.~6. The lower panel of the same figure shows the d$\chi$/dT vs. T plots to determine the transition temperatures more exactly. The peak in d$\chi$/dT occurs at 166~K for the cooling curve and at 209~K for the warming curve which shows that the hysteresis in the susceptibility is also about 40$-$45~K. It is interesting to note that there is a large diamagnetic drop (nearly 45\% reduction) in the susceptibility across the transition as we cool down from 300~K. Since the sample contains no magnetic atoms and the transition is not affected by magnetic field (both the resistivity and susceptibility transitions do not change with applied magnetic field as high as 8~Tesla), we estimate that the reduction in the Pauli susceptibility is almost 50\% across the transition. This indicates that the density of states (DOS) at the Fermi level could be changing condsiderably across this high temperature transition suggesting that an electronic transition accompanying or induced by the structural transition can not be ruled out. \\ 
Finally, the results of our differential scanning calorimetry measurements are shown in Fig.~7. Large peaks are seen in the data recorded while cooling and warming the sample. In the main panel of the figure, the peaks marked by arrows correspond to the anomalies which were seen in the resistivity and susceptibility measurements as well. However, an additional peak was observed at 135~K in the heating curve. The possibility of a matching second peak for the cooling scan could not be explored due to experimental limitations. It must be noted that no second peak was observed in either the cooling or warming scans in the resistivity or susceptibility measurements. These measurements were repeated three times and always revealed the same results.
The entropy associated with the transition (the peaks marked by the arrows) has been estimated and is shown in the inset of the figure~8. The entropy involved here is substantial but it is not the same for cooling and heating which is not understood at this juncture. It must be stressed here that it is not possible to separate the contribution of the structural phase transition from that arising from the electronic phase transition, which caused  the large drop in the susceptibility. More investigations, preferably on single crystal of Lu$_2$Ir$_3$Si$_5$ are needed for complete understanding of this(these) phase transition(s). Also, heat capacity measurements down to lower temperatures are required to understand the multiple peaks observed in the heating scan.

It is now worthwhile to compare the properties observed for Lu$_2$Ir$_3$Si$_5$ with those of the known CDW system Lu$_5$Ir$_4$Si$_{10}$. The Lu$_5$Ir$_4$Si$_{10}$ compound forms in the tetragonal Sc$_5$Co$_4$Si$_{10}$ type structure (P4/mbm) while Lu$_2$Ir$_3$Si$_5$ forms in the orthorhombic U$_2$Co$_3$Si$_5$ type structure (Ibam). Lu$_5$Ir$_4$Si$_{10}$ undergoes a transition below 83~K which has been shown to be a strongly coupled CDW ordering transition \cite{r14,r15}. The signatures of this transition in the bulk properties are (i) a step like upturn in the resistivity with $\Delta$$\rho$($T_{CDW})$/$\rho$(300K)~=~23\% and $\Delta$T~=~2~K. After the transition the resistivity still shows metallic behavior indicating only partial gapping of the Fermi surface due to the CDW, (ii) a diamagnetic drop in the magnetic susceptibility with jump size $\Delta$$\chi$~=~5$\times$10$^{-5}$ emu/mol with $\chi(300K)$~=~-6$\times$10$^{-5}$ and (iii) a huge spike of almost 100~J/mol (over the lattice contribution) in the heat capacity measurment. There is no structural transformation down to low temperatures and the compound becomes superconducting below 3.8~K. From the drop in the susceptibility and the heat capacity anomaly a 36\% reduction in the electronic density of states at the Fermi Surface has been estimated \cite{r17}. 

We have seen from our measurements on Lu$_2$Ir$_3$Si$_5$ that similar signatures are observed in the bulk properties for this compound. In particular (i) an upturn in the resistivity at 165~K with $\Delta$$\rho$($T_{CDW})$/$\rho$(300K)~$\approx$~35\% while cooling from 300~K and an upturn in the resistivity at 209~K with $\Delta$$\rho$($T_{CDW})$/$\rho$(300K)~$\approx$~22\% while warming to 300~K and the width of the transition for both warming and cooling cycles is $\approx$ 20~K. The resistivity remains metallic below this transition. (ii) a diamagnetic drop in the magnetic susceptibility with jump size $\Delta$$\chi$~=~3$\times$10$^{-4}$ emu/mol with $\chi(300K)$~=~-5$\times$10$^{-4}$ for both cooling (165~K) and warming (208~K) cycles and (iii) large peaks (20~J/mol while cooling and 30~J/mol while warming) in the specific heat. The structure of Lu$_2$Ir$_3$Si$_5$ changes below this transition and finally at low temperatures the compound undergoes a superconducting transition ($T_C\approx~3.5~K$). The structural change accompanying the electronic transition complicates the analysis of the electronic transition. Extra peaks in the specific heat also indicate more than one transition.\\ 
It is also worthwhile to note that the ionic size effect in the R$_5$Ir$_4$Si$_{10}$ compounds leads to the CDW transition occurring at higher temperatures for compounds with a larger unit cell volume \cite{r18}. For the R$_2$Ir$_3$Si$_5$ compounds, only Er$_2$Ir$_3$Si$_5$ shows a similar resistivte anomaly as the Lu$_2$Ir$_3$Si$_5$ compound \cite{r13}. However, it occurs at about 135~K and is much weaker. This indicates an ionic size effect opposite to that shown in the R$_5$Ir$_4$Si$_{10}$ compounds. Good quality samples of Tm$_2$Ir$_3$Si$_5$ and possibly Yb$_2$Ir$_3$Si$_5$ are required to be able to make a systematic ionic size effect analysis for the R$_2$Ir$_3$Si$_5$ compounds..

\section{CONCLUSION}
\label{sec:CON}
We have investigated the compound Lu$_2$Ir$_3$Si$_5$ using magnetic susceptibility, electrical resistivity, room and low temperature X-ray diffraction and Differential scanning Calorimetry (DSC) measurements. We find that it crystallizes in a U$_2$Co$_3$Si$_5$ type structure at room temperature and undergoes a structural and possibly an electronic transition below  150~K. It also undergoes
a superconducting transition below 3~K. It appears that Lu$_2$Ir$_3$Si$_5$ belongs to the growing group of CDW superconductors with 3-D structures such as, Lu$_5$Ir$_5$Si$_{10}$. However, unlike the latter, Lu$_2$Ir$_3$Si$_5$ exhibits a structural transition to another orthorhombic structure with a large unit cell. Studies on single crystal are essential to establish the nature of the high temperature transition.
\section{ACKNOWLEDGEMENTS}
\label{sec:ACK}
We would like to acknowledge Dr. N. P.Lalla of the Inter University Consortium, Indore, India for performing low temperature X-ray diffraction measurements.

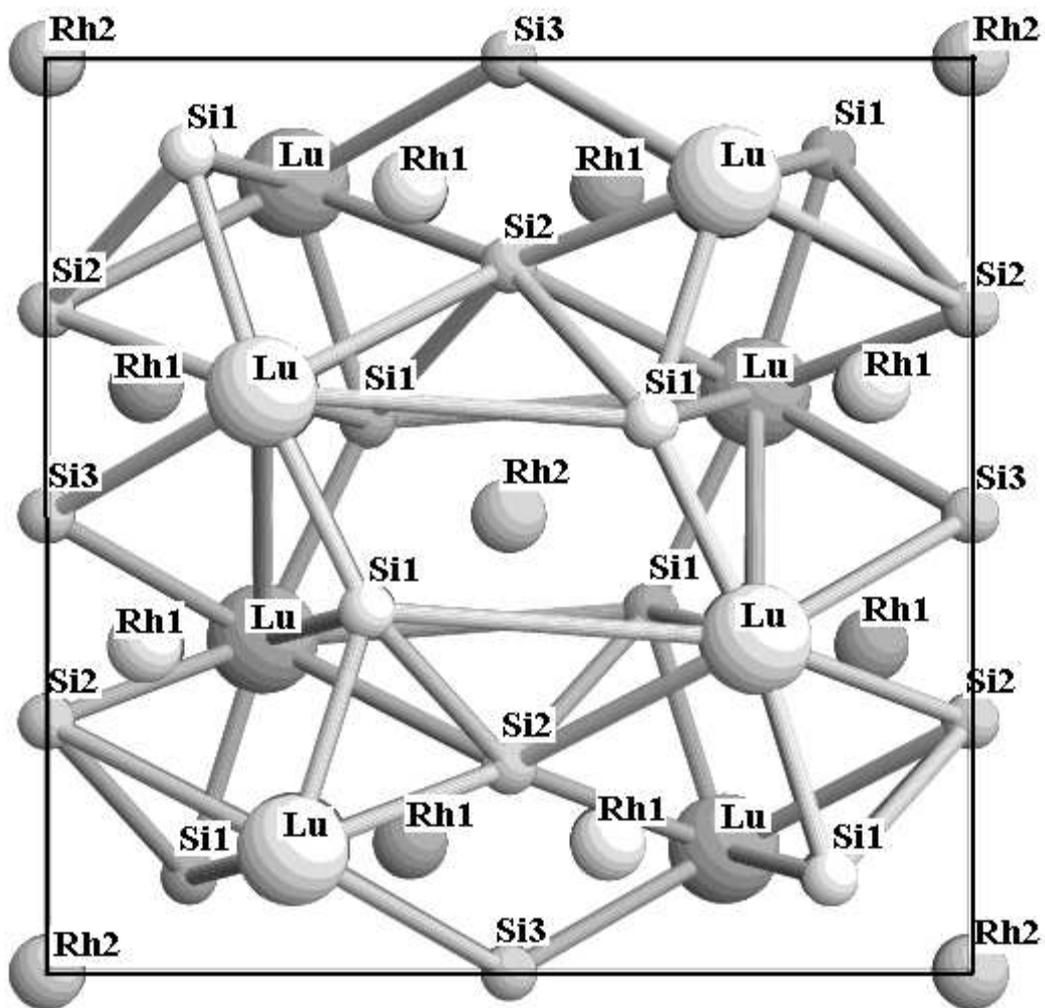
\begin{figure}
\caption{Structure of the unit cell of Lu$_2$Ir$_3$Si$_5$ which 
forms in the Orthorhombic U$_2$Co$_3$Si$_5$ type structure (space group
$Ibam$), as viewed along the c-axis.
\label{str}}
\end{figure}
\begin{figure}
\caption{Top panel (a) shows temperature dependence of the resistivity ($\rho$) of Lu$_2$Ir$_3$Si$_5$ taken while warming from 1.8 to 300~K. The inset of this panel shows the low 
temperature behavior of the resistivity on an expanded scale. The bottom panel (b) shows dc susceptibility from
1.8 to 10~K. The horizontal line is drawn where $\chi$=0 to emphasizes observation of 
the diamagnetic signal in the field-cooled state. 
\label{fig1}}
\end{figure}
\begin{figure}
\caption{Top panel (a) shows the resistivity ($\rho$) for temperature (T) scans while both cooling and warming  Lu$_2$Ir$_3$Si$_5$ between 100-300~K. The inset of the top panel shows the temperature dependence of d$\rho$/dT illustrating the hysteresis of the high temperature phase transition. The bottom panel (b) depicts the effect of thermal cycling on the resistivity for two cycles. The arrows indicate the start of cycle~1 and end of cycle~2.
\label{rhys}}
\end{figure}
\begin{figure}
\caption{Comparison of the powder X-ray diffraction data of Lu$_2$Ir$_3$Si$_5$ at 300~K and 87~K. 
\label{fcp1}}
\end{figure}
\begin{figure}
\caption{ The decay of resistance with time when the sample is left at 300~K
after being subjected to 7 thermal cycles.
\label{decay}}
\end{figure}
\begin{figure}
\caption{The top panel (a) shows the temperature (T) dependence of the susceptibility ($\chi$) of Lu$_2$Ir$_3$Si$_5$  while both cooling and warming 
between 85-300~K. The bottom panel (b) shows the d$\chi$/dT illustrating the hysteresis of the high temperature phase transition. 
\label{shys}}
\end{figure}
\begin{figure}
\caption{Plot of the heat-capacity (C$_p$) vs. temperature (T) of
Lu$_2$Ir$_3$Si$_5$ from 120 to 300~K while warming and cooling as measured by differential scanning calorimeter. The arrows mark the peaks which correspond to those which were also observed in the resistivity and susceptibility measurements. The inset shows the estimated entropy associated with the peaks marked by arrows in the main panel.
\label{fcp1}}
\end{figure}
\begin{table}
\caption{Unit cell parameters obtained from the {\bf FULL PROF} refinement of the room temperature powder X-ray diffraction data of Lu$_2$Ir$_3$Si$_5$ (Ibam). a~=~9.91457(5)~\AA ~, 
b~=~11.28665(5)~\AA~and c~=~5.72191(5)~\AA. Overall R factor 5.6.}
\begin{tabular}{cccccc}
atom&Ion&Wyck&x&y&z\\
 \tableline
Lu&76&8h& 0.26732&0.37011&0.00000\\
Ir&26&8h&0.10613&0.13415&0.00000\\
Ir&26&4d&0.50000&0.00000&0.25000\\
Si&14&4e&0.00000&0.00000&0.25000\\
Si&14&8h&0.00000&0.27396&0.25000\\
Si&14&8h&0.35730&0.09697&0.00000\\
\end{tabular}
\label{table 1}
\end{table}

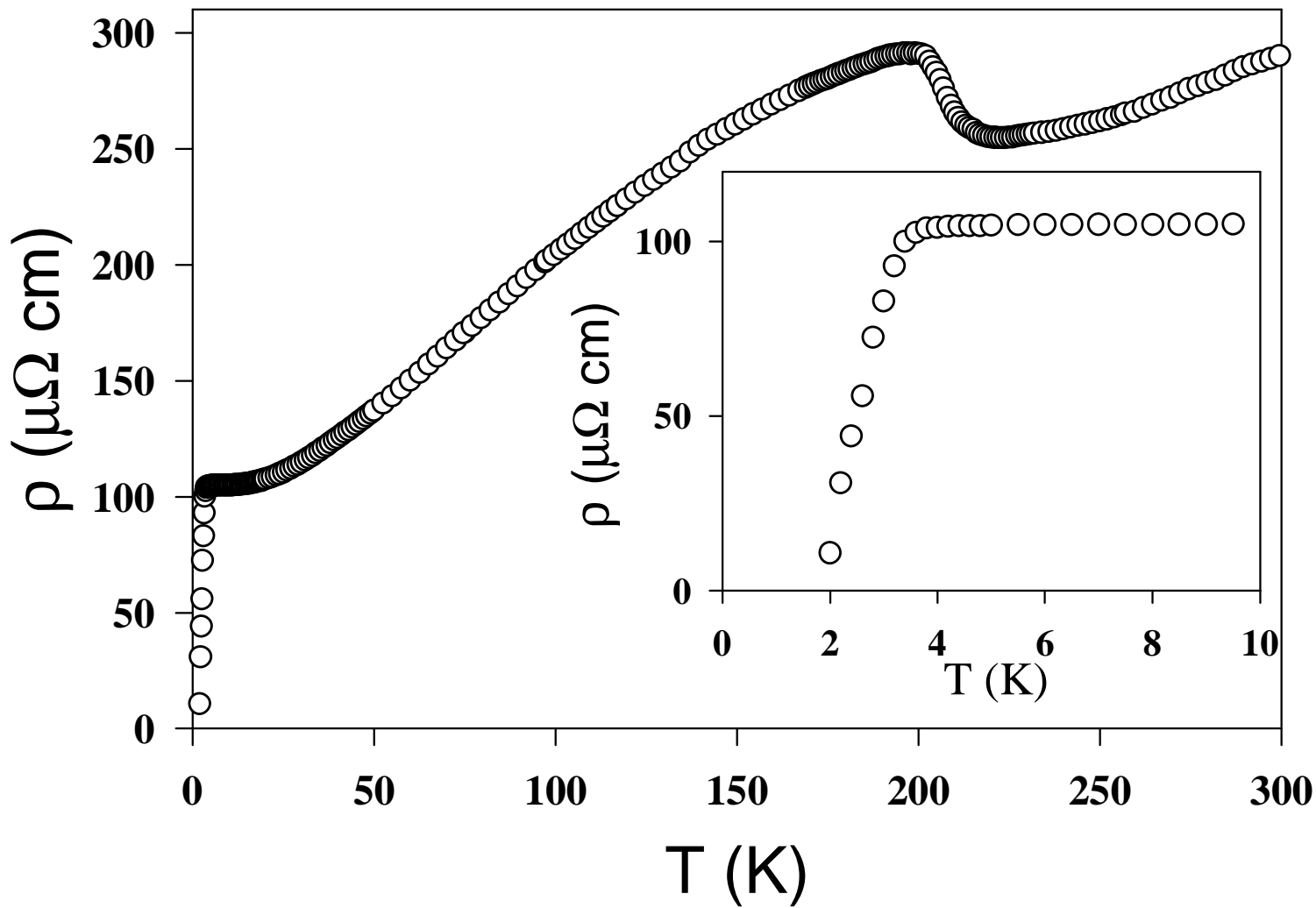
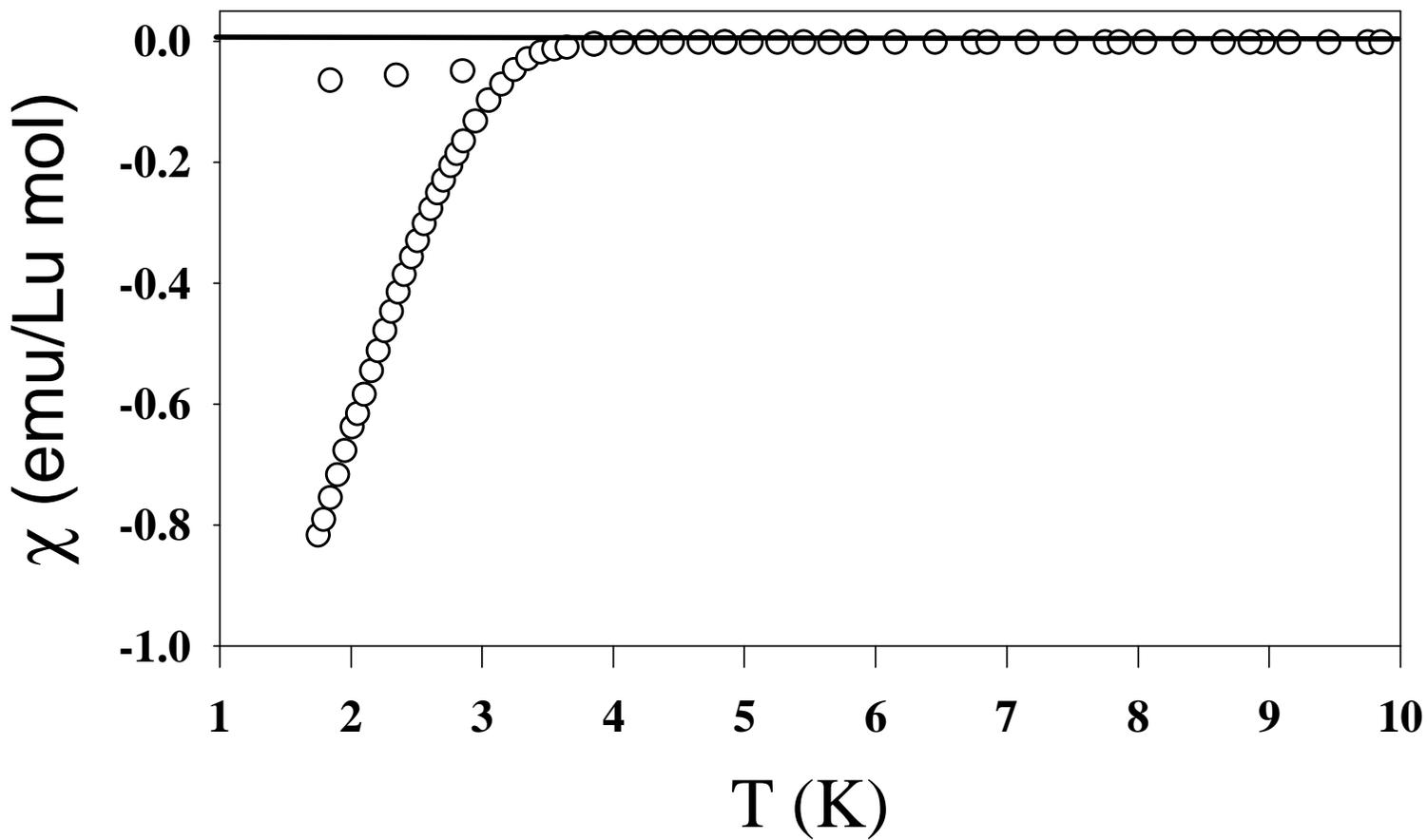
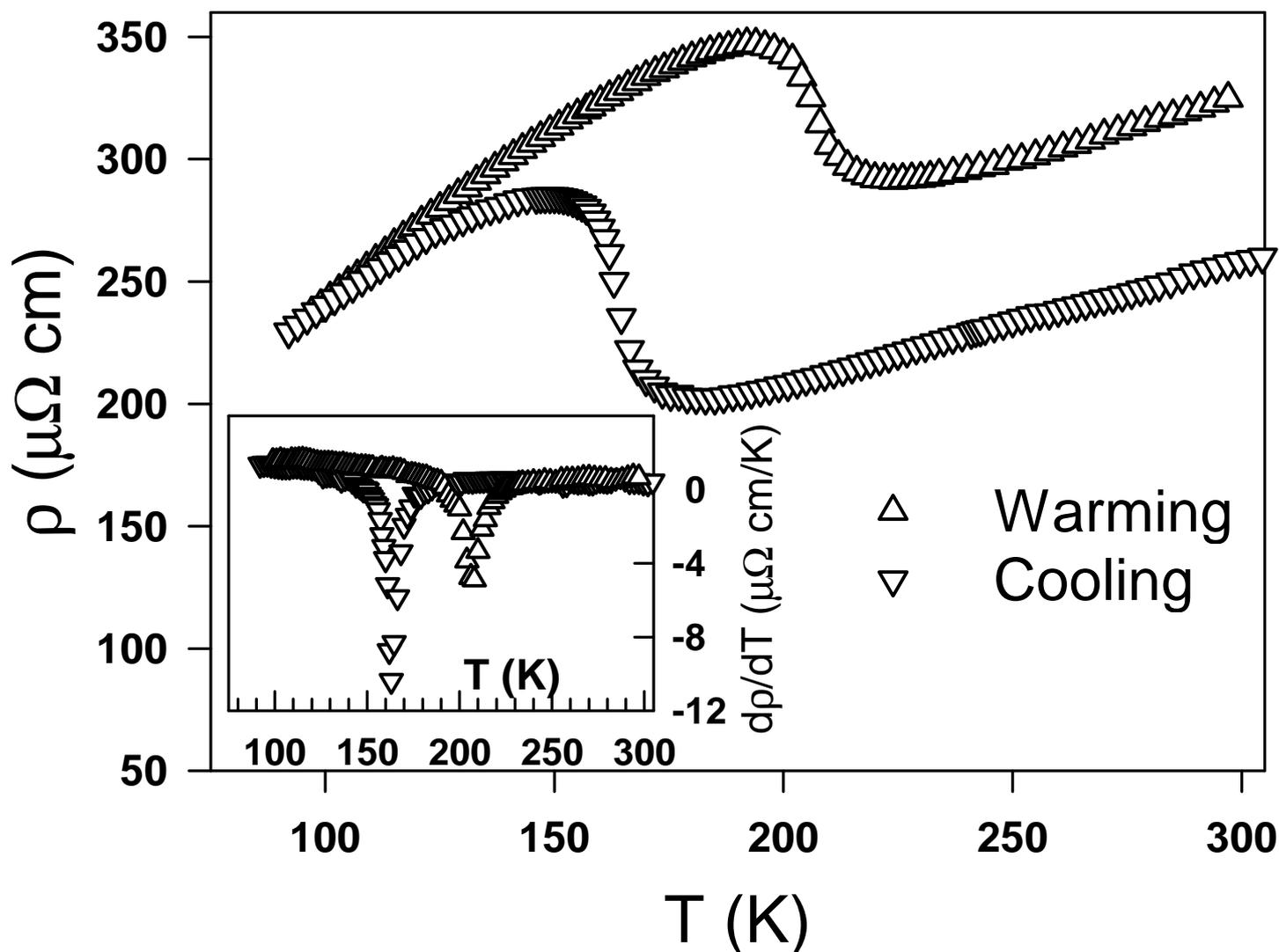
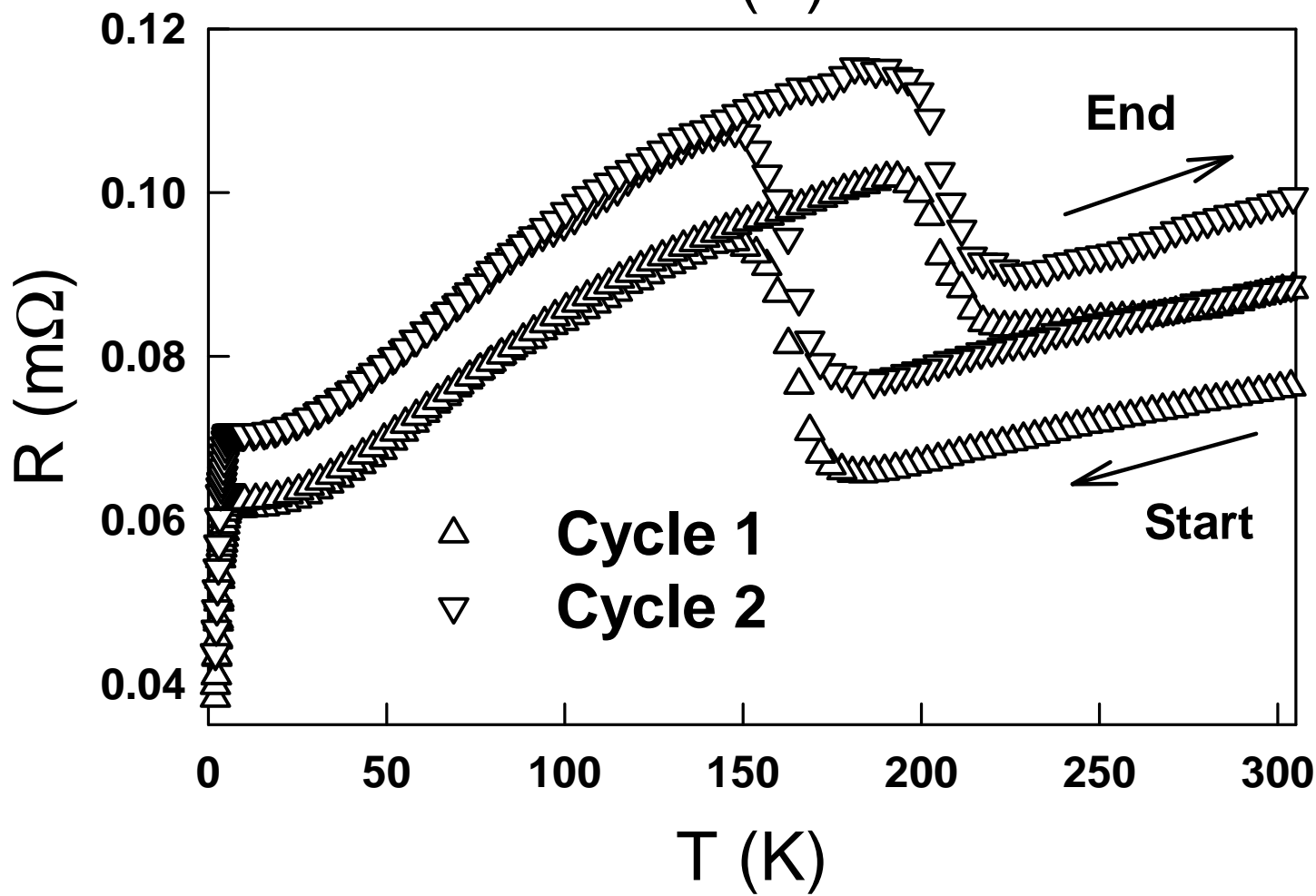
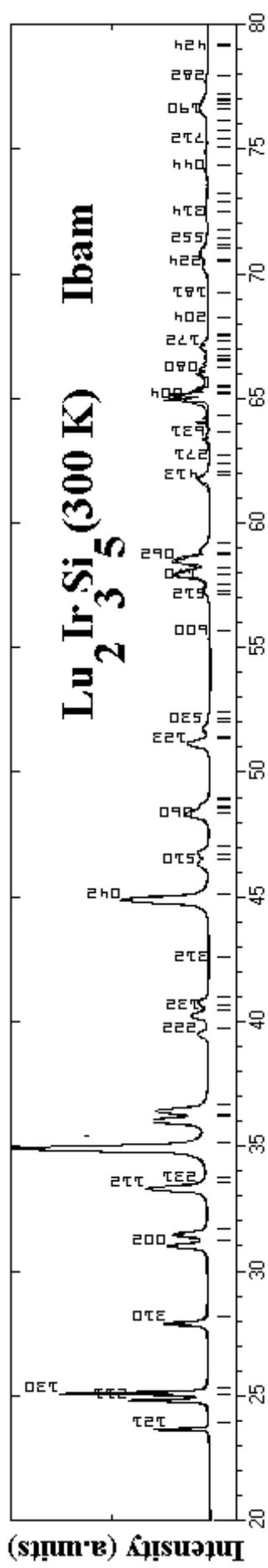
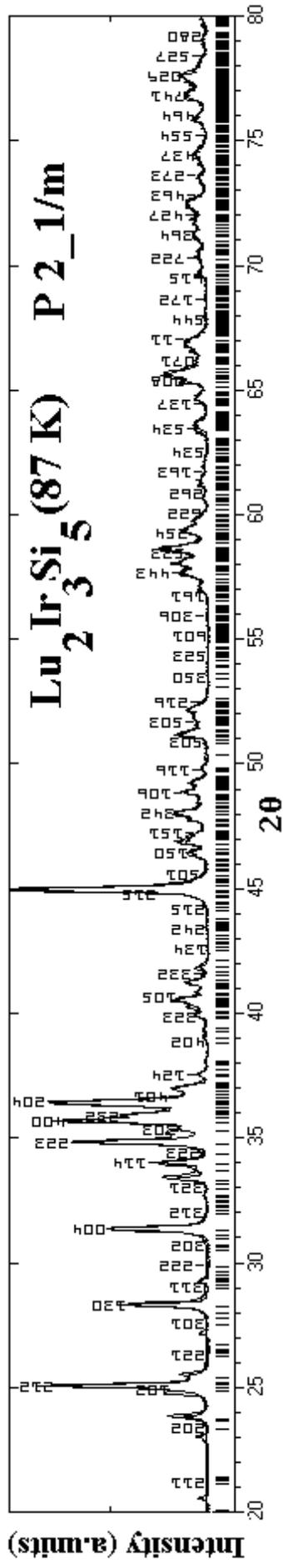
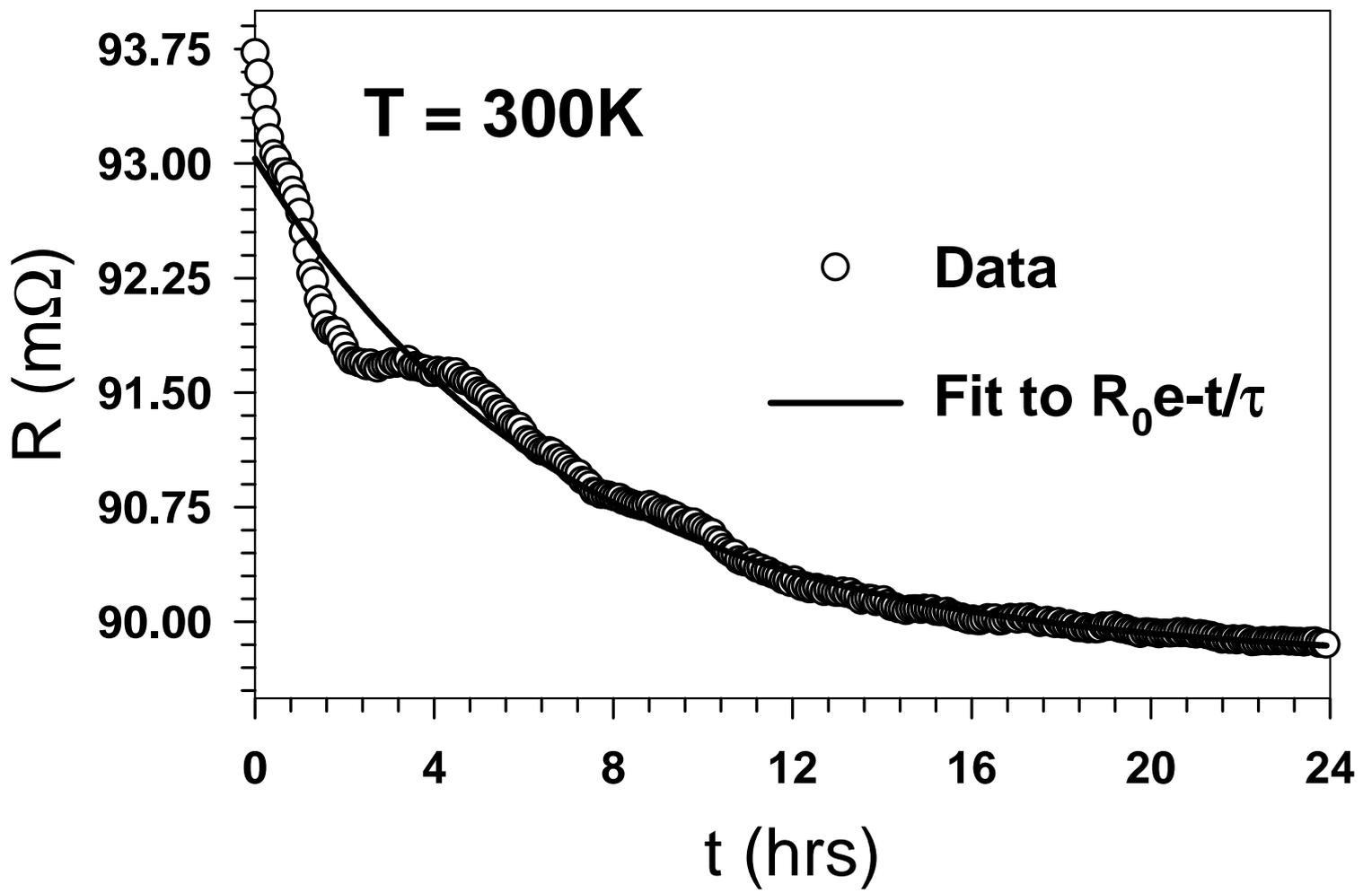
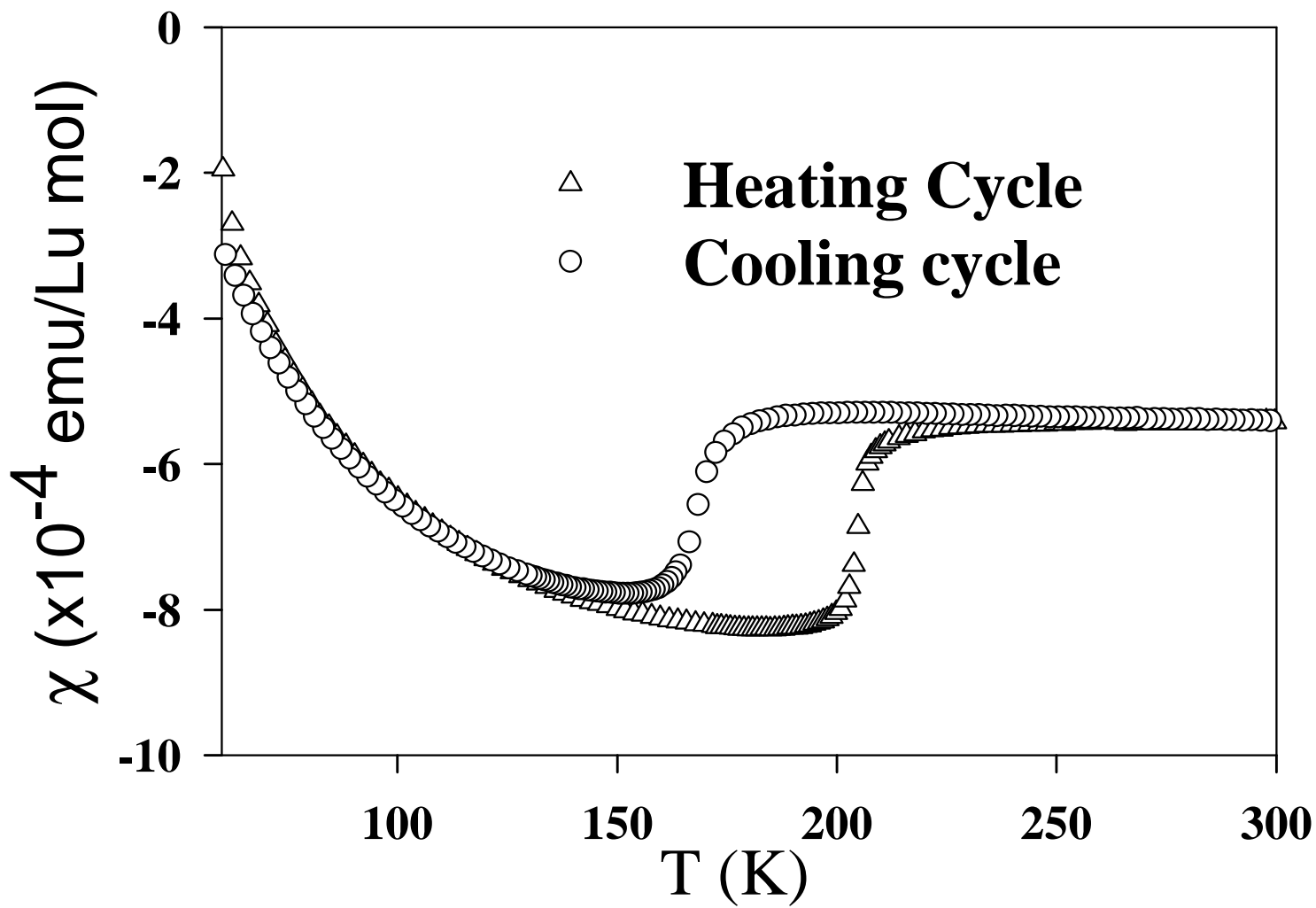
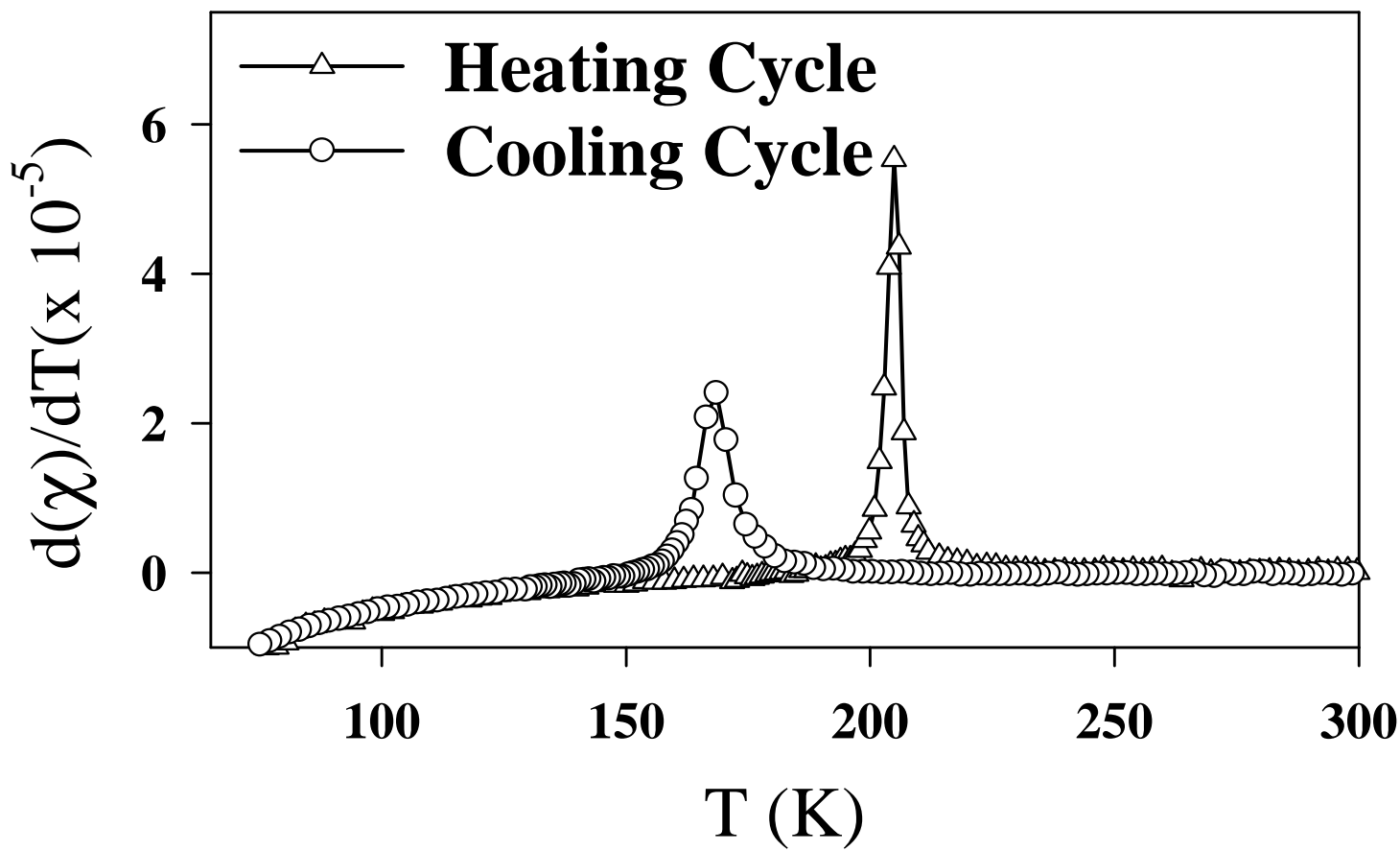
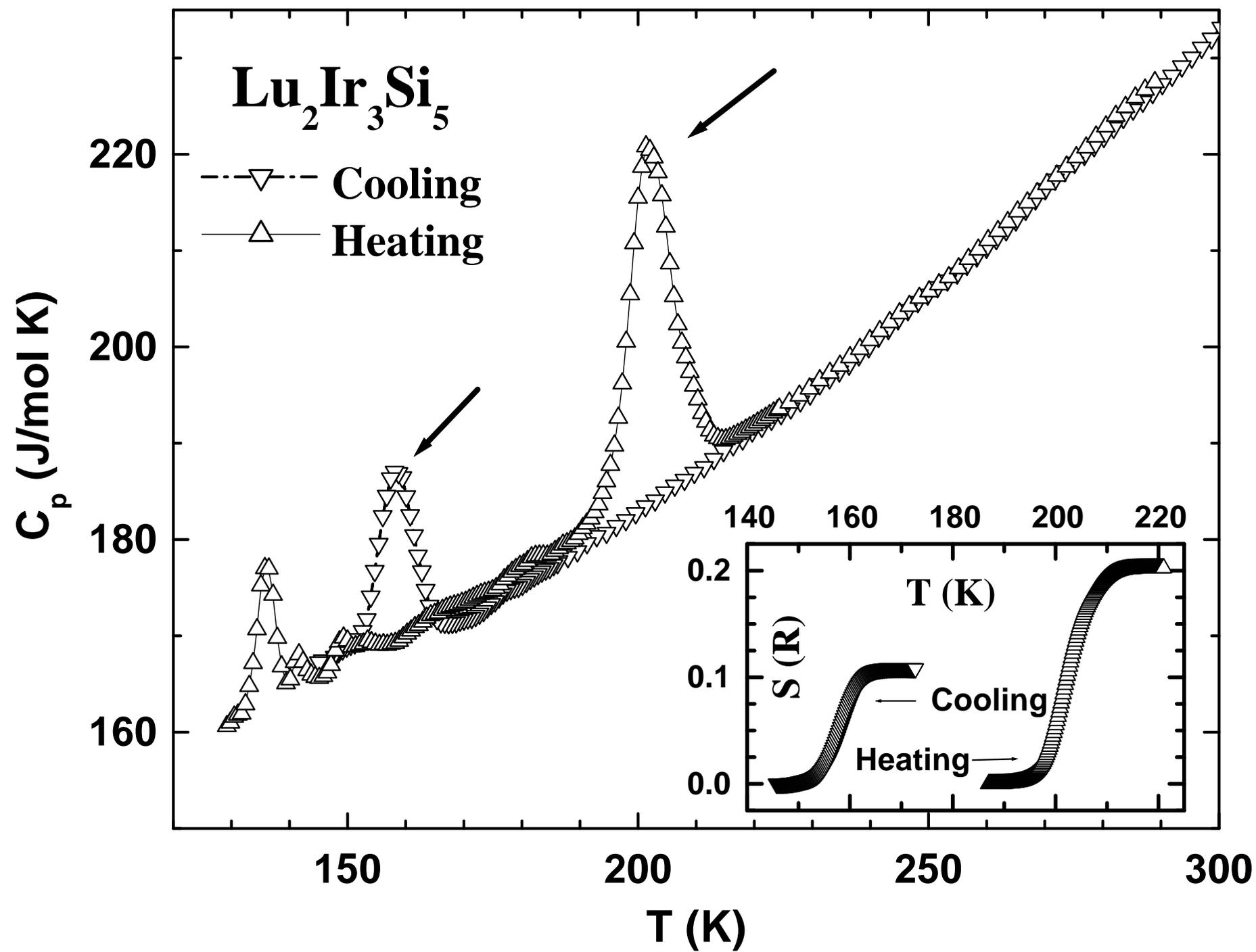
\end{document}